\documentclass[preprint,showpacs,showkeys,
preprintnumbers,amsmath,amssymb]{revtex4}
\usepackage{graphicx}
\usepackage{dcolumn}
\usepackage{bm}
\begin{document}
\newcommand{\be}{\begin{equation}}
\newcommand{\ee}{\end{equation}}
\newcommand{\der}[2]{{\frac{d #1}{d #2}}}
\newcommand{\beq}{\begin{eqnarray}}
\newcommand{\eeq}{\end{eqnarray}}
\newcommand{\pder}[2]{{\frac{{\partial} #1}{{\partial} #2}}}
\title{Geometric Phase in a Bose-Einstein Josephson Junction}

\author{Radha Balakrishnan}
\email{radha@imsc.res.in}
\author{Mitaxi Mehta}
\email{mitaxi@imsc.res.in}
\affiliation{ The Institute of Mathematical Sciences, 
C. I. T. Campus, Chennai 600 113, India}

\begin{abstract}
We calculate the geometric phase associated with  
the time evolution of the  wave
 function of a  Bose-Einstein condensate system in a
 double-well trap by using a model for tunneling between the wells.
 For a cyclic evolution, this phase is shown to
be  half the solid angle subtended
 by  the evolution of a unit vector whose $z$ component  
 and azimuthal angle are given, respectively, by the population difference and
 phase difference between the two condensates. For a non-cyclic evolution,
 an additional phase term arises.  We  show that the geometric
 phase can also be obtained  by mapping the tunneling equations
 on to the  equations of a space curve. The importance of a geometric phase
in the context of some recent experiments is pointed out.  
\end{abstract}

                            
\maketitle

\section{\label{sec:level1}Introduction}
 Bose-Einstein condensation (BEC) in a dilute gas of trapped ultracold
alkali atoms  has
been observed by several experimental groups \cite{Anderson}. This gives
rise to the possibility of understanding the nature of the condensate wave
function, and in particular, its phase \cite{Java}. 
 It is believed that the BEC phase transition occurs  due the breaking of
a global gauge symmetry of the Hamiltonian. Theoretically, a
BEC may be modeled by writing down the interacting many-body Hamiltonian in terms of
boson creation and annihilation operators 
$\Psi_{op} ^\dagger$ and $\Psi_{op}$. The order parameter
 is postulated to be the condensate wave function 
$\psi$=$<\Psi_{op}>$= $\rho e ^{i \theta}$, where $\rho$=$|\psi|^{2}$ is the
condensate density and $\theta$ is the phase of the wave function. The
Hamiltonian is gauge invariant, but the order parameter breaks this
symmetry. Using the  dynamical equation for  $\Psi_{op}$ found from the
Hamiltonian operator, the time evolution of  the condensate wave 
function $\psi$   can be shown to satisfy the following Gross-Pitaevskii
equation (GPE) \cite {gross}: 
$$
i \hbar \frac{\partial \psi} {\partial t} =
 - \frac{\hbar ^2}{2 m} \bigtriangledown^2 \psi + 
\left[ V_{ext}({\bf x}) + g_0 |\psi|^2 \right] \psi ,
\eqno(1.1)$$ 
where $V_{ext}$ is the external potential and $g_0= 4\pi
\hbar^2a/m$, $a$ and $m$ being the atomic scattering length and mass,
respectively. Although this equation has an underlying quantum
nature, the condensate has a macroscopic extent, suggesting the observation
 of quantum effects on a macroscopic scale.

In a striking experiment \cite{Andrews}, Andrews {\it et al} 
 have shown the existence of the macroscopic quantum
phase difference between two  bose condensates : They designed a double-well 
trap by using a laser
sheet to create a high barrier within a trapped condensate. On
{\it switching off} this barrier, the two condensates overlapped
to produce an interference pattern, showing phase coherence.
More interestingly, by {\it lowering} the laser sheet intensity,
 the barrier gets lowered, making it possible for the condensates
to {\it tunnel} through the barrier. Thus this double-well trap is
analogous to  a superconductor  Josephson junction \cite{barone},  
 and is referred to as the Boson Josephson junction (BJJ).
In an interesting paper, Smerzi {\it et al} \cite{Smerzi} have 
 set up the tunneling equations for the BJJ in a model. 
These are two coupled nonlinear ordinary
differential equations for the condensate wave functions in the
two wells. They have  studied the time  evolution of the
inter-well population difference and  phase  difference
 in this model, and predicted a novel `self-trapping' effect,
 i.e., the oscillation of the population difference  around a
 non-zero value,  for certain initial conditions and parameters.

 The tunneling dynamics  motivates the following question:
 Is there an underlying geometric phase associated with the time evolution of
the condensate wave function in a double-well trap ?  
As is well known by now, the concept of a geometric phase has
 been studied in various contexts, after it was introduced by  Berry \cite{Berry}
 in quantum mechanics. It had also been considered  much earlier by
Pancharatnam \cite{Panch} in the context of classical optics. 
Geometric phase and its various applications have been
studied intensively for over a decade now \cite{Wilc}.
Such a (non-integrable) phase arises when the time evolution of 
a system is such that the value of a variable    
 in a given state of the system depends on the path along
 which the state has been reached. In this paper, we calculate the geometric
 phase associated with the  time evolution of the BJJ wave function,
 for both cyclic and non-cyclic evolutions.  

 The plan of the paper is as follows: In Sec.II, we first review 
 Smerzi et al's  \cite{Smerzi} derivation of the tunneling equations.
 Keeping in mind that the geometric phase is gauge independent, we
 use a  certain gauge transformation to  reduce these
equations to a more convenient form.  
In Sec. III, we briefly outline the kinematic approach  formulated
  by Mukunda and Simon \cite{Mukunda} to define the geometric phase
 as applied to a two-level system.
 We then  solve  the BJJ tunneling equations, which are nonlinear
 differential equations, numerically by 
choosing some parameter values as
 an example. Using these solutions, we find the geometric phase 
explicitly, for both cyclic
 and non-cyclic evolution  of  the system. For a cyclic
 evolution, the phase difference and the population difference
 between the condensates in the two wells return to their
 original values. For this case,  
 the corresponding geometric  phase is 
  half the solid angle  generated by a {\it unit
 vector} whose $z$ component and azimuthal angle are given, 
respectively, by the population difference and the phase difference.
For non-cyclic evolution, an additional phase term is obtained. 
  In Sec. IV, we show that this geometric phase (for both types of evolution)
 can also be obtained by  first mapping the tunneling equations 
 to the equation for a unit
vector {\bf r}  and then identifying it with the tangent
 {\bf T} of a space curve. The space curve is  described using
 the so-called natural frame equations, which possess  an
underlying  natural gauge freedom. The unit triad of vectors
can be written down using the form of the condensate wave
functions in the two wells. The concept of Fermi-Walker
 parallel transport is then used to identify the  
 geometric phase.
   In  Sec.V, we employ  the usual Frenet frame to  obtain
 explicit expressions for the  curvature and torsion of the
space curve that gets  associated with the BJJ evolution. Section VI 
 contains a  summary and  discussion. 

\section{\label{sec:level2} The BJJ tunneling equations}
We begin by briefly describing the model used by Smerzi {\it et al}\cite{Smerzi}
to study the tunneling of the condensate between two wells.
 Let the total number of atoms in the double-well trap be $N$.
 Let $N_1$ and $N_2$ 
 denote the number of atoms in each well, such that $N_1+N_2= N$.
 To study the tunneling, the solution for the GPE (Eq. (1.1)) is assumed 
to be of the form 
$$
\psi = \psi_1(t) \Phi_1({\bf x}) + \psi_2(t) \Phi_2({\bf x}).
\eqno (2.1)$$
Here $\Phi_1, \Phi_2$ are the ground state solutions for the
isolated wells with $N_1=N_2=(N/2)$.
Using Eq. (2.1) in Eq. (1.1), one gets 
$$
i \hbar \pder{\psi_1} {t} = (E_1^0 + U_1 N_1) \psi_1 - V \psi_2
\eqno(2.2a)
$$
$$
i \hbar \pder{\psi_2} {t} = (E_2^0 + U_2 N_2) \psi_2 - V \psi_1.
\eqno(2.2b)
$$
The quantities $E^0_{1,2}, U_{1,2}$ and $V$ are constants 
obtained by taking the
 external field $V_{ext}({\bf x})$ in Eq. (1.1) to be independent of time:
$$
E^{0}_{1,2}=\int [(\hbar^{2}/2m)
  |\bigtriangledown  \Phi_{1,2}|^{2}
+\Phi^{2}_{1,2}~~ V_{ext}] ~ d^{3}x\nonumber
$$
$$
U_{1,2}=g_{0}\int\Phi^{4}_{1,2} d^{3}x,\nonumber
$$
$$
V=-\int[(\hbar^{2}/2m)(\bigtriangledown \Phi_{1}
  \bigtriangledown \Phi_{2})+\Phi_{1}\Phi_{2}~V_{ext}]~d^{3}x\nonumber
$$
 Further, $N_i=|\psi_i|^2$, for $i=1,2$.

 In this paper, we obtain the geometric phase associated with the
evolution in Eqs. (2.2). Recognizing that this phase is a gauge
 independent quantity, we find it convenient to apply the following gauge transformation 
 to the wave function  
$$
\left( \begin{array}{c}
    \psi_1 \\
    \psi_2 \end{array} \right) =
    \sqrt{N} e^{i \int{\eta(t') dt'}}
\left( \begin{array}{c}
    a \\
    b \end{array} \right). \label{eq.gauge}
\eqno(2.3)
$$ 
On substituting  Eq. (2.3)  in Eq. (2.2) and 
 setting $\hbar \eta(t) = -((E_1^0 + E_2^0) + N (U_1 |a|^2 
+ U_2 |b|^2))/2$,  the BJJ tunneling equations take on the
form 
$$
i \hbar \frac{d}{dt} \left( \begin{array}{c}
   a \\
   b \end{array} \right) =
  \left( \begin{array}{c c}
  \hbar \omega_0 & -V \\
   -V & -\hbar \omega_0  \end{array} \right)
   \left( \begin{array}{c}
   a \\
   b \end{array} \right) \nonumber \\ =
   M_{\omega_0} 
   \left( \begin{array}{c}
   a \\
   b \end{array} \right), \nonumber
\eqno(2.4)
$$
with
$$
\hbar \omega_0 = \frac{1}{2} (E_1^0 - E_2^0 +
 U_1 N_1 - U_2 N_2) 
\eqno(2.5) $$ 
 
  In Eq. (2.4),  $M_{\omega_{0}}$  denotes
 the (time-dependent) matrix on the right hand side. 
The advantage  of the gauge transformation
is that this matrix is traceless. Note also that 
since $N_1=N|a|^2$ and $N_2= N|b|^2$ appear (in $\omega_{0}$) in this matrix,  
Eqs. (2.4)  are two  coupled nonlinear differential equations.
From Eq. (2.3), we see that the  normalization condition
$ |\psi_1|^2 + |\psi_2|^2 = N$ implies $|a|^2 + |b|^2 = 1$.
 Thus without loss of generality, we write
$$ 
a~~=~~ \cos (\alpha/2)~~ e^{i\theta_1}~~;~~
b~~=~~ \sin (\alpha/2)~~ e^{i\theta_2} 
\eqno (2.6)$$
Let us denote  the difference in the population density of the two traps
 by  $z$ and the difference in the phases of the two condensates
 by $\phi$. From Eq. (2.6)  we thus have,
$$
z=(N_1-N_2)/N=(|a|^2-|b|^2)=\cos\alpha~~~;~~~\phi~=~(\theta_{2}-\theta_{1}).
\eqno (2.7)$$ 
 By suitably combining Eq. (2.4) and its complex conjugate,
 and using Eq. (2.7),  the
 nonlinear coupled equations for $z$ and $\phi$ are  found to be (on setting $\hbar=1$)  
$$
\der{z}{t} = - V \sqrt{1-z^2} \sin{\phi}\eqno(2.8a)
$$
$$
\der{\phi}{t} = \Lambda z +  V \frac {z}{\sqrt{1-z^2}}
\cos{\phi} + \Delta E.
\eqno(2.8b)
$$ 
Here we have defined \cite{Smerzi} 
$$
\Delta E =
(E_1^0 -E_2^0)/2 + N (U_1-U_2)/4~~;~~
\Lambda = N (U_1 + U_2)/4 
\eqno(2.9)$$ 
and the time has been
reparametrized as $t \rightarrow 2 t$.
It is interesting to note that the above equations
 can also be written as  Hamilton's equations,
by treating $z$ and $\phi$ as the canonically
conjugate variables.  The  classical Hamiltonian is easily verified to be
$$
H_{cl} = \Lambda \frac{z^2}{2} - V \sqrt{1 - z^2} \cos{\phi}
 + \Delta E z.
 \label{eq.Hcl}
\eqno(2.10) $$ 
 This describes a non-rigid pendulum with a length proportional to
$\sqrt {1-z^{2}}$, which decreases with the ``momentum'' $z$.
We  write down  the expression
for $\omega_0$  appearing in Eq. (2.5) in the form    
$$
\omega_0 = \Delta E + \Lambda z, 
\eqno(2.11)$$
by using Eq. (2.9).
Finally, setting $z=\cos \alpha$ in Eqs. (2.8a) and (2.8b), we obtain
$$
\der{\alpha}{t} = V \sin{\phi}
\eqno(2.12a)
$$ 
$$
\der{\phi}{t} = \Lambda \cos{\alpha} +  V \cot{\alpha} 
\cos{\phi} + \Delta E \eqno(2.12b)
$$

Equations (2.8), or equivalently, Eqs. (2.12), represent the tunneling equations. 
For convenience, $V$ can be absorbed in time $t$ and all energies 
can be measured in units of $V$. Note that $\Delta E$ is the asymmetry 
between the two wells, as seen from Eq. (2.9).
We consider two special limits:

(1) Interacting Bose system in a  symmetric trap: 
 $\Lambda \ne 0$, $\Delta E = 0$.\\
 From Eq. (2.9),  $E^{0}_{1}=E^{0}_{2}$ and $U_1 = U_2=U$, and $\Lambda=UN_T$ is nonzero.
Thus Eq. (2.10) becomes
$$
H_{cl} = \Lambda \frac{z^2}{2} - V \sqrt{1 - z^2} \cos{\phi}
\eqno(2.13) $$ 

 In Figs. 1, 2  and 3,  we have obtained
 the $(z,\phi)$ phase portraits for this case, by using Eq. (2.13).
Let $\Lambda$  be replaced by the dimensionless quantity $(\Lambda/V)$.
 For $\Lambda < 1$ there exist
periodic oscillations around the zero-state $(0,0)$ and the non-
trapped $\pi$ state $(0,\pi)$. There are no rotational states. 
(see Fig. 1 for $\Lambda =0.5$). For $\Lambda > 1$ two new trapped 
$\pi$-states appear 
at $(z^*,\pi)$ and $(-z^*, \pi)$ with $z^* = \sqrt{\Lambda^{2}-1}/\Lambda$.
The traped $\pi$-states are clearly visible in Fig. 2 ($\Lambda = 1.3$). 
 As $\Lambda$ increases, $z^* \rightarrow 1 $.
 For $\Lambda > 2$, rotational states also appear as seen in
 Fig. 3 ($\Lambda=5$).  \\

(2)  Non-interacting Bose system : $\Lambda = 0$. \\
For an ideal bose gas, the interactions $U_1$ and $U_2$ are negligible, and Eq. (2.9)
yields $\Lambda=0$.
In this limit the kinetic energy term of the Hamiltonian $H_{cl}$ (2.10)
vanishes and hence the non-rigid pendulum analogy is not valid 
anymore. (However, we remark that the tunneling Hamiltonian coincides with that of
a two-component BEC in the rotating frame approximation. This will be discussed
in the last section.)
We get
$$
H_{cl} = -V \sqrt{1-z^2} \cos{\phi} + \Delta E z
\eqno (2.14) $$
In Figs. 4, 5 and 6, we have obtained the phase space portraits for this case.
Again, as in case (1), the energy is measured in units of $V$.  
When $\Delta E = 0$, we have a symmetric trap. There
are oscillations around the $(0,0)$-state and the non-trapped $(0,\pi)$-state.
There are no rotational orbits, as seen in Fig. 4.  With increase in $\Delta E$,
 rotational orbits appear, which explore the full range of $\phi$. This is shown 
in Figs. 5 and 6. The oscillations are now around $(z=-z^{*},\phi=0)$ 
and $(z=+z^{*},\phi=\pm\pi)$. That is, oscillations around  $\phi=0$  shifts towards 
$z = -1$ which is energetically more stable, while  the $\phi=\pi$ fixed point 
moves towards $z = 1$ which is a local
energy maximum. 
Note that the increase in $\Delta E$ causes  a  non-trapped $\pi$ state
to   become a trapped $\pi$ state with oscillations around
a non-zero population difference $z^*$. The new fixed point is given
by $z^{*} = \sqrt{((\Delta E/V)^2/(1 + (\Delta E/V)^2))}$ which tends to 1
as $\Delta E/V$ goes to infinity. 

We now proceed to  show how the geometric phase associated with the
BJJ dynamics can be computed.

\section{\label{sec:level3}Geometric Phase using the kinematic approach}

 In this section we derive the expression for the geometric phase
for the BJJ evolution using the kinematic approach developed by 
Mukunda and Simon \cite{Mukunda}. We first briefly outline the
basic ideas. One starts  with a complex Hilbert space and
considers a  subset ${\cal N}_0$  made up of  {\it unit}
 complex  vectors denoted by $\psi(t)$.
Let  ${\cal C}_0$ be a smooth one parameter curve, consisting of a family
of vectors $\psi(t)$, with $0<t<T$. Since $\psi(t)$ are
 unit vectors for all values of $t$, 
$$
Re(\psi(t),\dot{\psi}(t)) = 0, 
$$
giving 
$$
(\psi(t),\dot{\psi}(t)) = i \mbox{Im}(\psi(t),\dot{\psi}(t)) 
\eqno(3.1) $$  
where the dot denotes the derivative with respect to $t$.
Under a gauge transformation defined by a real function 
$\alpha(t)$, ${\cal C}_0 \rightarrow {\cal C}_0'$ and $\psi(t)
\rightarrow \psi'(t) = e^{i\alpha(t)} \psi(t)$, $t \in (0,T)$,
one finds,
$$
\mbox{Im}(\psi'(t), \dot{\psi}'(t)) = \mbox{Im}(\psi(t), \dot{\psi}(t)) + 
\der{\alpha}{t}
\eqno(3.2) $$
On the other hand, we have
$$
arg(\psi'(0),\psi'(T)) - arg(\psi(0),\psi(T))=\int_0^T (\frac{d\alpha}{dt})~dt
$$ 
Integrating Eq. (3.2) and using the above equation, we see that
a gauge invariant function,
$$
arg(\psi'(0),\psi'(T)) - \mbox{Im} \int_0^T dt~~(\psi'(t),\dot{\psi'}(t))
\eqno(3.3) $$
can be constructed. Next one introduces the space ${\cal R}_0$ of unit rays,
which is the quotient of ${\cal N}_0$ with respect to the $U(1)$ action
$\psi \rightarrow e^{i \alpha} \psi$. It can then be shown that
the curves ${\cal C}_0 $ and ${\cal C}'_0$ (obtained by a gauge transformation)
in ${\cal N}_0$ both project to the same curve 
${c}_0$ in the ray space ${\cal R}_0$.
Thus the gauge invariance of the functional (3.3) implies that it is
a functional only of the image ${c}_0$ of ${\cal C}_0 $. This functional
can be shown to be reparametrisation invariant as the integrand is
linear in $\dot{\psi}'(t)$. One thus defines the 
geometric phase $\phi_g(c_{0})$ as the following gauge and 
reparametrisation invariant functional,
$$
\phi_g[{c}_0] =  
 arg(\psi'(0),\psi'(T)) - \mbox{Im} \int_0^T dt~~(\psi'(t),\dot{\psi}'(t))
\eqno(3.4) $$
Identifying the first term  in Eq. (3.4) as the total phase
$$
\phi_{p}=  arg(\psi'(0),\psi'(T))\eqno(3.4a) 
$$
and the second term in Eq. (3.4)  as the dynamical phase
$$
\phi_{d}= \mbox{Im} \int_0^T dt~~(\psi'(t),\dot{\psi}'(t)), \eqno(3.4b)
$$
we get 
$$
\phi_g[{c}_0] = \phi_p[{\cal C}_0 ] - \phi_d[{\cal C}_0],
\eqno(3.5) $$

 Now for a given ${c}_0 \in {\cal R}_0$ 
the gauge freedom in the choice of 
${\cal C}_0 \in {\cal N}_0$ can be used to 
express $\phi_g[{c}_0]$ in different
but equivalent ways. The Pancharatnam connection takes $\psi(0)$
and $\psi(T)$ to be in phase, giving $\phi_p = 0$, leading to
$\phi_g = - \phi_d$. Another prescription takes a
lift of ${\cal C}_0$ so that the integrand of $\phi_{d}$ in (3.4b) vanishes. This leads 
to $\phi_g =  \phi_p$.

 Let us now understand what these two prescriptions imply for 
the BJJ evolution equations. Using equation (2.6) the family
of unit vectors is given by,
$$
\psi' = 
\left( \begin{array}{c}
    a \\
    b \end{array} \right) =
     e^{i \theta_1(t)}
\left( \begin{array}{c}
    \cos (\alpha(t)/2) \\
    \sin (\alpha(t)/2) e^{i \phi(t)} \end{array} \right)=e^{i\theta_{1}(t)}\psi, 
\eqno(3.6) $$  
where $\phi = (\theta_2 - \theta_1)$. Clearly, the angle $\theta_1$ in
Eq. (3.6) represents the gauge freedom in the wave function $\psi'(t)$. The
prescriptions essentially fix the gauge $\theta_{1}$. Using Eq. (3.6) in Eq. (3.4a),
 a short calculation leads to  the following total phase:
$$
\phi_p = arg(\psi'(0), \psi'(T))
       = (\theta_1(T) - \theta_1(0)) + \Delta.
\eqno(3.7) $$
Here,
\begin{widetext}
$$
\Delta = \tan^{-1} 
\left[ \frac{\sin (\alpha(0)/2) \sin (\alpha(T)/2) \sin(\phi(T) - \phi(0))}
  {\cos (\alpha(0)/2) \cos (\alpha(T)/2) +
   \sin (\alpha(0)/2) \sin (\alpha(T)/2) \cos(\phi(T) - \phi(0))} \right]
\eqno(3.8) $$
\end{widetext}
The integrand of the dynamical phase $\phi_{d}$ can be calculated using Eq. (3.6)
 in Eq. (3.4b) to give 
$$
\mbox{Im}(\psi', \der{\psi'}{t}) = \dot{\theta_1} + \sin^2 (\alpha/2) \dot{\phi}.
\eqno(3.9) $$ 
Thus
$$
\phi_{d}= (\theta_{1}(T)-\theta_{1}(0))+~\int_0^T \sin^2 (\alpha/2) \dot{\phi}~dt.
\eqno(3.9a)$$
From Eqs. (3.7) and (3.9a), we get the geometric phase to be
$$
\phi_{g}= \phi_{p}-\phi_{d}= -~\int_0^T \sin^2 (\alpha/2) \dot{\phi}~dt + 
\Delta.\eqno(3.9b)
$$
Next, we also calculate the geometric phase expression for
two different gauge prescriptions, for the sake of completeness.

{\it The Pancharatnam Prescription}.

 Here $\phi_p = 0$, thus Eq. (3.7) yields,
$$
\theta_1(T) - \theta_1(0) = -\Delta.
\eqno(3.10) $$
Given the end-points $\alpha(0), \alpha(T), \phi(0), \phi(T)$
on the curve ${\cal C}_0$ one can thus determine 
$\Delta$ from Eq. (3.8), and hence the gauge $\theta_1(T) - \theta_1(0)$.
From Eq. (3.9),
$$
\phi_d = \int \mbox{Im}(\psi, \dot{\psi}) dt \\
       = \theta_1(T) - \theta_1(0) + \int{\sin^2 (\alpha/2) 
\dot{\phi}} dt.
\eqno(3.11) $$
 
Substituting Eq. (3.10) in Eq. (3.11),
$$
\phi_d = - \Delta + \int_0^T{\sin^2 (\alpha/2)~~ \dot\phi} dt.
\eqno(3.12) $$
Since $\phi_g$=-$\phi_d$, we get 
$$
\phi_{g}=  -~\int_0^T \sin^2 (\alpha/2) \dot{\phi}~dt + \Delta,
\eqno(3.13) $$
which agrees with Eq. (3.9b).

{\it The Horizontal Lift}.

Here the integrand in the $\phi_d$ definition is zero, giving
$$
\dot{\theta_1} = -\sin^2 {\alpha/2}~~ \dot{\phi},
$$
so that the gauge is determined in this case to be   
$$
\theta_1(T) - \theta_1(0)=  -\int_0^T \sin^2{(\alpha/2)}~~ \dot{\phi}~dt.
\eqno(3.14) $$
Substituting this in Eq. (3.7),
$$
\phi_p = -\int_0^T (\sin^2{\alpha/2}) ~~\dot{\phi}~dt + \Delta. \eqno(3.15)
$$
Since $\phi_d = 0$ here, we get an expression for $\phi_g$ which is the same as
Eq. (3.13). Thus  the two prescriptions give the same geometric phase,
 which  holds for both cyclic and noncyclic evolutions. 

For a cyclic evolution, it is clear from (3.8) that
$\Delta = 0$. Hence the geometric phase is just  minus half the 
solid angle $\Omega$ subtended by the closed curve generated  on a sphere by  the
 tip of  a  unit vector $\bf {r}$:   
$$
\bf {r} = (\sin \alpha \cos \phi, \sin \alpha \sin \phi, \cos \alpha)
\eqno(3.16) $$
 Here, $\alpha$ and $\phi$ denote the polar and azimuthal
angles of $\bf {r}$.

\noindent {\it An Example}:\\
As an example  we consider  an interacting Bose system with$\Lambda=5$, 
 in a symmetric trap, i.e.,  $\Delta E = 0$. 
 As already mentioned, the phase space portrait for this case is  
found from Eq. (2.13) and is given in Fig. 3.
The value of $\Lambda$ selected is quite generic because further
increase in $\Lambda$ does not change the character of the phase-
space portrait much, apart from moving the $\pi$-state towards $z=1$.
Since $\cos \alpha=z$, the geometric phase $\phi_g$ given in 
Eq. (3.9b) can be re-expressed as
$$
\phi_{g}=  \frac{1}{2}~\int_0^T (z-1) \dot{\phi}~dt + \Delta.
\eqno(3.17) $$
While comparing the geometric phase from the above equation to
that in Eq. (3.4), it is necessary to keep in mind that in the Hamiltonian
formalism the time has been scaled and so one has to use the appropriate
value of time in Eq. (3.17).
We solve Eqs. (2.8a) and (2.8b) numerically for $(z(t),\phi(t))$, for a  given 
 initial condition $(z(0),\phi(0))$ at time $t=0$.
Using these solutions, we find the  solutions for the corresponding 
unit vectors $\bf {r}$
given in Eq. (3.16). These yield the the path on the unit sphere plotted in Fig. 7.
Next, we  substitute the solution  for $(z(t),\phi(t))$ in Eq. (3.17), 
to find $\phi_g(t)$ numerically, for various times $t$,
in the range $0\le t\le T$, where $T$ denotes the full period of the orbit concerned.  
Our plot for the time dependence  of $\phi_{g}(t)$ over a period for a librational
 orbit is given in Fig. 8, while  Fig. 9 gives that plot for  a rotational orbit. 

 We conclude this section with the following remark. There exists an
interesting geometrical representation of a two level system in terms
of the time evolution of a unit vector $\bf {r}$.
 In the next section, we identify ${\bf r}$ with the
tangent of a space curve, and provide a classical
differential geometric approach to derive the geometric
phase $\phi_g$ associated with the   
 BJJ evolution.    

\section{\label{sec:level4}Geometric Phase using   space curve approach}

In this section, we derive the geometric phase associated with
the Bose condensate tunneling dynamics by providing a geometric
visualisation of this two level system. Firstly it is possible to
show\cite{Feyn} that the tunneling equations (2.4) for the two-level wave function,
$$
\psi = \left( \begin{array}{c} 
       a \\ 
       b \end{array} \right)   
     = e^{i \theta} 
        \left( \begin{array}{c} 
       \cos (\alpha/2) \\
       \sin (\alpha/2) e^{i \phi} \end{array} \right),
\eqno(4.1) $$
with $\phi = \theta_2 - \theta_1$ can be mapped to the following vector
evolution equation.
$$
d {\bf r} / dt = \mbox{\boldmath $\omega$ }  \times   {\bf r}.
\eqno(4.2) $$ 

Here, in  cartesian coordinates ,
$$
\mbox{\boldmath $\omega$} = (-2 V, 0, 2 \omega_0)
\eqno(4.3) $$ 

$$ 
{\bf r} = (a^*b + a b^*, i(ab^* - a^*b), |a|^2 - |b|^2).
\eqno(4.4)$$  

 Using the definitions of $a$ and $b$ given in Eq. (2.6), Eq. (4.4) is
readily seen to be identical to the unit vector $\bf {r}$
in Eq. (3.16). 

 Urbantke \cite{Urbantke} has shown that corresponding to a wave function
of the form (4.1), in addition to ${\bf r}$ two more unit vectors
${\bf P'}$, ${\bf Q'}$ can be defined, such that the set $({\bf r},   
{\bf P'},{\bf Q'})$ forms a unit orthogonal right-handed triad. This 
is achieved by defining a complex vector $\bf {Z'}$ as follows :
$$
{\bf Z'} = {\bf P'} + i {\bf Q'} = ( (a^2-b^2), i(a^2+b^2), -2a b).
 \eqno(4.5)
$$
On using Eq. (2.6) in Eq. (4.5), we get,
\begin{widetext}
$$
{\bf Z'} = e^{2 i \theta_1} (\cos^2(\alpha/2) - \sin^2(\alpha/2) e^{2 i \phi}, 
        i (cos^2 (\alpha/2) + \sin^2(\alpha/2) e^{2 i \phi}),
        - \sin(\alpha) e^{i \phi})
\eqno(4.6)
$$
\end{widetext}
It can be easily verified that,
$$
|{\bf r}| = |{\bf P'}| = |{\bf Q'}| = 1 ~~;~~ 
 {\bf r} \cdot {\bf P'} = {\bf r} \cdot {\bf Q'} = {\bf P'} \cdot {\bf Q'} = 0
\eqno(4.7)
$$

Clearly now as ${\bf r}$ evolves with time, 
so does the $({\bf P'},{\bf Q'})$ plane. The {\it total phase} 
$\Gamma_p$ accumulated by ${\bf Z'}(t)$ in
time $T$ is given by,
$$
\Gamma_p = arg({\bf Z'}(0)^* \cdot {\bf Z'}(T)) 
\eqno(4.8)
$$
Substituting for ${\bf Z'}$  from Eq. (4.6) into Eq. (4.8),
after some algebra we obtain,
$$
\Gamma_p = 2 [(\theta_1(T)-\theta_1(0)) + \Delta] 
\eqno(4.9)
$$
where $\Delta$ is identical to the expression Eq. (3.8) obtained
in the kinematic approach in Sec. III. Thus from Eqs. (3.7) and (4.9) we get 
$$
\Gamma_p = 2 \phi_p
\eqno(4.10)
$$
$\phi_p$ being the total phase of $\psi$ in the kinematic approach.

First we find the {\it total phase} rotation $\gamma_{p}$ 
associated  with the rotation of ${\bf P'}$ or $({\bf Q'})$
as follows. It is defined by 
$$
\cos \gamma_p ={\bf P'}(T)\cdot{\bf P'}(0)={\bf Q'}(T)\cdot{\bf Q'}(0).
$$
Further, it  is easy to see  geometrically that 
${\bf P'}(T)\cdot{\bf Q'}(0)=-{\bf Q'}(T)\cdot{\bf P'}(0)= \sin\gamma_p$. 
Substituting ${\bf Z'}={\bf P'}+i{\bf Q'}$ in Eq. (4.8) and using the above relations, 
we  can show that the total phase  
$$ 
\gamma_p= -\Gamma_p= -2[(\theta_{1}(T)-\theta_{1}(0))+\Delta],
\eqno(4.11)$$
where we have used Eq. (4.9).

 Next we wish to find the {\it dynamical phase} $\gamma_d$ associated with 
$({\bf P'},{\bf Q'})$ rotation, which is induced by the specific dynamical equations
of the frame  $({\bf r},{\bf P'},{\bf Q'})$. 
This is a little more involved, and we proceed as follows.

From Eq. (4.6), we have 
$$
{\bf Z'}=e^{2i\theta_1}{\bf Z} 
\eqno(4.12)$$
This immediately leads to 
$$
{\bf P'}+i{\bf Q'}=e^{2i\theta_1}({\bf P}+i{\bf Q}),
\eqno(4.13)$$
Comparing this with Eq. (4.6) yields 
\begin{widetext}
$$
{\bf P}=  (\cos^2(\alpha/2) - \sin^2(\alpha/2) \cos{2  \phi}, 
        - \sin^2 (\alpha/2) \sin{2  \phi},
        - \sin\alpha \cos{ \phi})
\eqno(4.14a)$$
$$
{\bf Q}=  ( - \sin^2(\alpha/2) \sin {2  \phi}, 
         cos^2 (\alpha/2) + \sin^2(\alpha/2) \cos{2  \phi},
        - \sin\alpha \sin{ \phi})
\eqno(4.14b)$$
\end{widetext}
It can be easily verified that $({\bf r}, {\bf P},{\bf Q})$ is also a right-handed triad.

A short calculation using Eqs. (3.16)   and (4.14) shows that we can write 
$$
d {\bf r} / dt = X {\bf P}+ Y {\bf Q},
\eqno(4.15) $$ 
where 
$$
X=(\frac{d\alpha}{dt}) \cos \phi-(\sin\alpha \frac{d\phi}{dt})\sin \phi
\eqno(4.16a)$$
$$
Y= (\sin\alpha \frac{d\phi}{dt})\cos\phi+(\frac{d\alpha}{dt})\sin\phi 
\eqno(4.16b)$$
 
Obviously, there is a gauge freedom $2\theta_1$ in the choice of $({\bf P},{\bf Q})$.
We immediately see this from Eq. (4.13): 
$$
{\bf P'}={\bf P} \cos \beta -{\bf Q} \sin \beta
\eqno(4.17a)$$
$$
{\bf Q'}= {\bf P} \sin \beta + {\bf Q} \cos \beta,
\eqno(4.17b)$$ 
where   
$$
\beta=2\theta_{1}
\eqno(4.18)$$
represents the gauge freedom.
Using Eqs. (4.17), we solve for $({\bf P},{\bf Q})$ in terms of $({\bf P'},{\bf Q'})$.
Substituting them in Eq. (4.15) yields 

$$
\frac{d{\bf r}}{dt} = \alpha_1 {\bf P'} + \alpha_2 {\bf Q'}
\eqno(4.19)
$$
where
$$
\alpha_1= \frac{d\alpha}{dt} \cos(\phi+\beta) -
(\sin\alpha \frac{d\phi}{dt}) \sin(\phi+\beta) 
\eqno(4.20a) 
$$
$$
\alpha_2 = \frac{d\alpha}{dt} \sin(\phi+\beta)+ 
(\sin\alpha \frac{d\phi}{dt}) \cos(\phi+\beta) 
\eqno(4.20b) 
$$
Since  $({\bf r},{\bf P'},{\bf Q'})$ is an orthonormal triad, Eq.
(4.19) immediately implies,
$$
\frac{d{\bf P'}}{dt} = -\alpha_1 {\bf r} + \alpha_3 {\bf Q'}
  \eqno(4.21)
$$
$$ 
\frac{d{\bf Q'}}{dt} = -\alpha_2 {\bf r} - \alpha_3 {\bf P'}
  \eqno(4.22)
$$
where $\alpha_3$ is to be determined. In the space curve language,
if ${\bf r}$ is identified  with the tangent ${\bf T}$, then $\alpha_1$
and $\alpha_2$ are the components of the curvature vector 
$\frac{d{\bf T}}{dt}$ along ${\bf P'}$ and ${\bf Q'}$ respectively.
 Further, equations (4.19), (4.21) and (4.22) describe
 the equations for a 
space curve in a ``natural frame''
 $({\bf T},{\bf P'},{\bf Q'})$. We
remark that the Frenet frame\cite{vais} corresponds to $\alpha_2 = 0$,
${\bf P'}$ is the normal ${\bf n}$, ${\bf Q'}$  is the binormal
 ${\bf b}$. Further, $\alpha_3$  is the  torsion $\tau$ and 
$\alpha_1$ is the curvature $K$.
On setting  $\alpha_2$ = $0$, we get from Eq. (4.20), the
following ``Frenet gauge''  $\beta_{F}$:  
$$
\tan(\beta_F + \phi) = \sin\alpha \frac{d\phi}{dt} / (\frac{d\alpha}{dt})
  \eqno(4.23)
$$

 Working with the natural frame, a short calculation using Eqs. 
(4.15) to (4.18) yields,
$$
\alpha_3 = {\bf T} \cdot (\dot{\bf T} \times \ddot{\bf T})/|\dot{\bf T}|^2 -
           \frac{d}{dt} \tan^{-1}(\frac{\alpha_2}{\alpha_1}).
  \eqno(4.24)
  \eqno(4.24)
$$

Next using the cartesian representation of ${\bf T} = {\bf r}$
given in Eq. (4.4), a lengthy but straightforward calculation leads 
to,
$$
{\bf T} \cdot (\dot{\bf T} \times \ddot{\bf T})/|\dot{\bf T}|^2 =
\cos\alpha \frac{d\phi}{dt} + \frac{d}{dt} 
 \tan^{-1} \left[ \frac{\sin\alpha \frac{d\phi}{dt}} 
 {\frac{d\alpha}{dt}} \right].
 \eqno(4.25)
$$

Substituting Eq. (4.25) and (4.20) in Eq. (4.24) and using the
formula $\tan^{-1}A - \tan^{-1} B = \tan^{-1}((A - B)/(1 + AB))$, we
obtain,
$$
\alpha_3 = \cos{\alpha} \frac{d\phi}{dt} - \frac{d(\phi+\beta)}{dt}
=-2\sin^{2}\frac{\alpha}{2}\frac{d\phi}{dt}-\frac{d\beta}{dt} 
\eqno(4.26) $$
Note that the time derivative of the gauge freedom $\beta(t)$
appears in $\alpha_3$. 

 We write Eqs. (4.19), (4.21) and (4.22) in a compact form,
$$
\frac{d{\bf T}}{dt} = \mbox{\boldmath $\xi$} \times {\bf T};~~
\frac{d{\bf P'}}{dt} =\mbox {\boldmath $\xi$} \times {\bf P'};~~
\frac{d{\bf Q'}}{dt} =\mbox {\boldmath $\xi$} \times {\bf Q'}.
 \eqno(4.27)
$$
Here  $\mbox{\boldmath $\xi$}$ is given by,
$$
\mbox{\boldmath $\xi$} = \alpha_3 {\bf T} + 
\alpha_1 {\bf Q'} - \alpha_2 {\bf P'}. \eqno(4.28)
$$

Eqs. (4.27) show that the natural frame 
$({\bf T},{\bf P'},{\bf Q'})$
rotates with an angular velocity $\mbox{\boldmath $\xi$}$, as it moves along the
space curve. As is obvious, $\alpha_1$ and $\alpha_2$ are components
of $\mbox{\boldmath {$\xi$}}$ along the ${\bf Q'}$ and ${\bf P'}$ axes
 respectively
and hence {\it tilt} the $({\bf P'},{\bf Q'})$ plane. On the other hand,
$\alpha_3$  merely {\it rotates} this plane around ${\bf T}$. Thus in time $T$,
the $({\bf P'},{\bf Q'})$ plane gets rotated by an angle $\gamma_d =
\int_0^T \alpha_3 dt$. Such a frame is defined  using Fermi-Walker parallel transport as \cite{Synge}, 
$$\frac{DA^i}{dt} = 
  \{ (\alpha_1 {\bf Q'} - \alpha_2 {\bf P'}) \times A \}^i.
 $$
 Using the expression for $\alpha_3$ given in (4.26) we obtain the dynamical phase $\gamma_{d}$
 associated with $({\bf P'},{\bf Q'})$  plane to be 
$$
\gamma_d = \int_0^T \alpha_3 dt = 
   -2\int_0^T (\sin^{2}\frac{\alpha}{2})~~ \frac{d\phi}{dt}~~dt -2(\theta_{1}(T)-\theta_{1}(0)), 
\eqno(4.29)
$$
since $\beta=2\theta_{1}$, from Eq. (4.18). 

Subtracting Eq. (4.29) from the  expression for the total phase $\gamma_p$ given in Eq. (4.11)
 we obtain  the geometric phase $\gamma_g$ associated with $({\bf P'},{\bf Q'})$ rotation to be 
$$
\gamma_g = \gamma_p-\gamma_d= 2[\int_0^{T}(\sin^{2}\frac{\alpha}{2})
\frac{d\phi}{dt}~~dt-\Delta]\eqno(4.30)$$
Note that the term involving the  gauge freedom $\beta$ cancels out here too, as in the kinematic
approach. Comparing Eq. (4.30) with Eq. (3.9b), we see that 
$$
\gamma_g=-2\phi_g
$$
In other words, the geometric phase $\phi_g$ 
associated with the wave function is minus half of that associated with the $({\bf P'},{\bf Q'})$ rotation.  
For a cyclic evolution, $\Delta=0$. Here, on computing $\gamma_g$,
the geometric phase $\phi_g$ becomes just minus half the solid angle, as we saw in Sec. III.
In summary, by mapping the evolution equation for
the wavefunction to the dynamical equation for an orthonormal
triad $({\bf T},{\bf P'},{\bf Q'})$ and identifying the triad to be
a natural frame on a space curve, enables us to provide a  
purely geometrical visualisation of the geometric phase
of a two level system.

 Our general result is valid for any two level system with the
wavefunction (4.1), since we did not use the specific BJJ equations
(4.2) and (4.3) in its derivation. By finding the solutions $\alpha(t)$
and $\phi(t)$ for the nonlinear equations (2.12) numerically for 
given initial conditions, $\gamma_g$ can be computed and 
is exactly -$2 \phi_g$, with $\phi_g$ values as plotted
 in Figs. 8 and 9.
 
\section{\label{sec:level5}Geometric parameters associated with BJJ dynamics}

In the last section, we discussed the mapping of the BJJ
tunneling equations to a space curve which is described using
 equations
for a ``natural frame''. This description involves three geometrical
parameters $\alpha_i$ which are shown to depend on a gauge parameter
$\beta$ (see Eq. (4.20) and (4.24)).

 The usual description of a space curve is in terms of a Frenet
frame\cite{vais}, with the curvature $K$ and torsion $\tau$ as the geometric
parameters. As explained in Sec. IV, working with the Frenet
frame implies {\it fixing}  $\beta = \beta_F$, defined in Eq. (4.19).
In this section we work with the Frenet frame to determine the geometric parameters $K$ and
$\tau$ of the space curve associated with the BJJ dynamics, in
terms of the physical parameters $V, \Delta E$ and $\Lambda$
and discuss certain special cases of interest.

 As mentioned in Sec. IV, in the Frenet frame, $\alpha_1 = K$,
$\alpha_2 = 0, ~~ \alpha_3 = \tau, ~~ {\bf P} = {\bf n}$ and
${\bf Q} = {\bf b}$ in Eqs. (4.15) to (4.18). In this frame, we  have the usual 
Frenet-Serret equations\cite{vais},
$$
\der{{\bf T}}{t} = K {\bf n}~~,~~ \der{{\bf n}}{t} = -K {\bf T} + \tau {\bf b}~~;~~  \der{{\bf b}}{t} = -\tau {\bf n}  
\eqno(5.1a) 
$$
Thus,
$$
\der{{\bf T}}{t} = \mbox{\boldmath $\xi$}_F \times {\bf T}
~~;~~ \der{{\bf n}}{t} =\mbox {\boldmath $\xi$}_F \times {\bf n}
~~;~~  \der{{\bf b}}{t} = \mbox{\boldmath $\xi$}_F \times {\bf b}.
\eqno(5.1b) $$
Also, 
$$
K^2 = (\der{{\bf T}}{t})^2 = ~~\sin^2\alpha (\der{\phi}{t})^2 +
     (\der{\alpha}{t})^2
\eqno(5.2) $$
and 
$$
\tau = {\bf T} \cdot (\dot{\bf T} \times \ddot{\bf T})/K^2\linebreak 
     = \cos \alpha \der{\phi}{t} + \der{}{t} \tan^{-1}
      ( \frac{\sin \alpha \der{\phi}{t}}{\der{\alpha}{t}}),
\eqno(5.3) $$
where the cartesian representation (3.16) , ${\bf T} = {\bf r}$
has been used. On using Eqs. (2.12) for $\der{\alpha}{t},
~~\der{\phi}{t}$ in Eqs. (5.2) and (5.3) respectively,  we get
$$
K = 2 (V^2 + \hbar^2 \omega_0^2 \sin^2 \alpha - 
 V^2 \sin^2 \alpha \cos^2 \phi\linebreak
$$
$$
 + 2 V \hbar \omega_0 \cos \alpha
 \sin \alpha \cos \phi)^{1/2},
\eqno(5.4)
$$   
and

$$
\tau = \cos\alpha(\hbar \omega_0~~ +~~ V \cot \alpha \cos \phi)\nonumber
$$
$$ 
+\der{}{t} \tan^{-1}(\frac{\hbar \omega_0 \sin \alpha 
+ V \cos \alpha \cos \phi}{V \sin \phi}).\eqno(5.5)
$$ 
Equations (5.4) and (5.5) give the curvature and torsion of the
space curve created by the BJJ dynamics Eq. (2.4), with parameters $V$ and
$\hbar \omega_0$. Here $\alpha$ and $\phi$ are solutions of the 
integrable equations (2.12).

 Since ${\bf r}$ is identified with ${\bf T}$, we also have,
for the BJJ system,
$$
\der{{\bf T}}{t} = \mbox {\boldmath {$\omega$}} \times {\bf T}, ~~ 
\mbox{\boldmath {$\omega$}} = (-2V,0,2 \omega_0),
 \eqno(5.6) $$
From Eq. (4.2) therefore $K^2$ can also be written as,
$$
K^2 = (\der{{\bf T}}{t})^2 =
 (\mbox{\boldmath {$\omega$}})^2 -(\mbox {\boldmath
 {$\omega$}} \cdot {\bf T})^2.
 \eqno(5.7) $$ 

Using the definition of the matrix $M_{\omega_0}$ given in
equation (2.4), a simple calculation shows that
$2<M_{\omega_0}> = {\bf \omega} \cdot {\bf T}$,
  yielding 

$$ 
4 <M_{\omega_0}^2> = (\mbox {\boldmath {$\omega$}})^2.
\eqno(5.8) $$ 

Using Eq. (5.8) in Eq. (5.7), we get 
$$
K = 2 ( <M_{\omega_0}^2> -  <M_{\omega_0}>^2)^{1/2}.
\eqno(5.9) $$ 

Now from the first equation in Eq. (5.1) it is clear that the 
distance traveled by the tip of ${\bf T}$ on the unit sphere in 
time $dt$ is $ds = K dt$.This is the well known \cite{AP}
Fubini-Study metric. Thus we see that the curvature $K$ which determines
the geometric quantity $ds$ is given by the variance of the tunneling
matrix $M_{\omega_0}$ (see Eq. (2.4)) for a two level system.
 As seen from the equation
(5.3), the torsion integral $\int \tau dt$ measures the anholonomy of
the frame, i.e. a path dependent geometric quantity given by the solid
angle associated with a cyclic evolution of ${\bf T}$.

Recalling that the population density difference between the two
traps is given by $z$ and the phase difference by $\phi$, it is 
instructive to write the geometric quantities $K$ and $\tau$ in 
terms of these physical quantities and the system parameters
$V, \Delta E$ and $\Lambda$: from Eq. (5.4),
$$
K = 2 (V^2 + (\Delta E + \Lambda z)^2 (1-z^2) -
 V^2 (1-z^2) \cos^2 \phi \nonumber
$$
$$
+ 2 V (\Delta E + \Lambda z) z
 \sqrt{1-z^2} \cos \phi)^{1/2}.
\eqno(5.10) $$
After a short calculation $K$ can be written as,
$$
K =  2 [V^2 + (\Delta E + \Lambda z)^2 - (H_{cl}+
 \frac{\Lambda z^{2}}{2})^{2}]^{1/2},
\eqno(5.11) $$
where $H_{cl}$ is the effective classical Hamiltonian given in Eq.
(2.10), which leads to the integrable dynamics of $z$ and $\phi$.
Next from Eq. (5.5) we obtain $\tau$:
$$
\tau = z (\Delta E + \Lambda z + \frac {V z}{\sqrt{1-z^2}} \cos{\phi})
\nonumber
$$
$$ 
+\der{}{t} \tan^{-1} \frac{(\Delta E + \Lambda z) \sqrt{1-z^2} +
  V z \cos{\phi}} {V \sin {\phi}}.
\eqno(5.12) $$
Using the expression for $H_{cl}$ once again, we get  
$$
\tau = H_{cl} + \frac{\Lambda z^{2}}{2} + 
   \frac{V}{\sqrt{1-z^2}} \cos{\phi} \nonumber
$$
$$
  +\der{}{t} \tan^{-1} \frac{(\Delta E + \Lambda z) \sqrt{1-z^2} +
  V z \cos{\phi}} {V \sin {\phi}}.
\eqno(5.13) $$

We consider some special cases:\\
{\bf(1)}  Interacting Bose system with no external potential: ( $V=0$,  $\Lambda \ne 0$.)\\
 From Eq. (2.8), setting $V=0$, we get $z$=constant.
 This in turn yields $\tau = H_{cl} + \Lambda z^2/2 $=constant and
 $K = 2 (\Delta E + \Lambda z) \sqrt{1-z^2}$ =constant.
i.e., the underlying geometry is that of a circular helix with a constant pitch. \\
{\bf(2)} The Ideal Bose Gas in an external potential: ($\Lambda=0$, $V \ne 0$.)\\
 If one considers a non-interacting Bose system then $\Lambda = 0$ (see (2.9).
In this limit the kinetic energy term of the Hamiltonian $H_{cl}$ (2.10)
vanishes and hence the non-rigid pendulum analogy is not valid 
anymore. However the tunneling Hamiltonian coincides with that of
a two-component BEC in the rotation frame approximation \cite{Will}.
Here,
$$
H_{cl} = -V \sqrt{1-z^2} \cos{\phi} + \Delta E z
\eqno (5.14) $$
From Eq. (5.13), we see that,
$$
\tau = H_{cl} + \frac{V}{\sqrt{1-z^2}} \cos{\phi} +
  \der{}{t} \tan^{-1} \frac{\Delta E  \sqrt{1-z^2} +
  V z \cos{\phi}} {V \sin {\phi}}.
\eqno(5.15) $$
 Further, from Eq. (5.11),
$$
K = 2 (V^2 + (\Delta E)^2 - H_{cl})^{1/2}.
\eqno(5.16) $$
Since $H_{cl}$ is a constant under time evolution, Eq. (5.16) shows
that the curvature $K$ is a constant. However, the torsion $\tau$
is time-dependent in this case. 
Since $K$ is a constant, the path length on
the unit sphere as given by the Fubini-Study metric is linearly
dependent on time for this case. (If $V$ and $\Delta E$ are
made time dependent, then $K$ is not a constant any more.)\\
{\bf(3)} The linear limit:\\
For a symmetric trap with $\Delta E = 0$ in the small oscillations
limit, linearizing Eqs. (2.8) in  {\it both} $z$ and $\phi$, for $|z| << 1$, $|\phi| << 1$, we get,
$$
\der{z}{t} = -V \phi,~~ \der{\phi}{t} = (\Lambda + V) z,
\eqno(5.17) $$
and
$$
H = (\Lambda + V) \frac{z^2}{2} + V \frac{\phi^2}{2} 
$$
This is just the harmonic oscillator limit and analytical solutions
are known. The corresponding expressions for $K$ and $\tau$ can
be calculated using Eqs (5.11) and (5.12).

{\bf(4)} The pendulum limit: \\
For a symmetric trap with $\Delta E=0$, linearizing Eqs. (2.8) in $z$ only, 
with  $\Lambda >> 1$ we get the equations of a pendulum,
$$
\der{z}{t} = -V \sin{\phi}~~ \der{\phi}{t} = \Lambda z
\eqno(5.18) $$
As is well known the solutions for $z$ can be written in terms of Elliptic
functions thus $K$ and $\tau$  can be obtained from 
 Eq.s (5.11) and (5.12).

Finally, for a symmetric trap $\Delta E=0$, with no linearizing 
approximations,  though the analytical solution of Eqs. (2.8) 
can be found, it is easier to work with numerical solutions instead,
using which $K$ and $\tau$ can be computed numerically using the 
expressions (5.11) and (5.12). 

\section{\label{sec:level6}Summary of results and discussion}
The geometric phase associated with the time evolution
of the wave function of a Bose-Einstein condensate in a double well
trap  has been found using a quantum approach as in Sec. III. 
We have explicitly computed the geometric phase $\phi_{g}$
 for both cyclic and noncyclic evolutions of the 
condensate population density difference $z$ and phase difference
$\phi$ in the two wells, by taking an example.
In Sec IV, we have shown that
the geometric phase can also be derived using a  classical
differential geometric approach, by essentially mapping the evolution
of the two states to a framed space curve with natural moving frames 
along the curve. The unit tangent vector ${\bf T}$ to the curve 
has $\alpha=\cos^{-1}z$ as the polar angle and $\phi$ as the azimuthal angle. 
As we have shown, here the geometric phase arises due to the path-dependent
rotation of the frame perpendicular to {\bf T} as the system evolves in time.   

 In an experimental set up,   
suppose one designs  a double-well trap by creating a barrier within a trapped
condensate with $N$ atoms, using a laser sheet. At time $t=0$, let $N_{1}(0)$ and $N_{2}(0)$
represent the number of condensate atoms in the two neighbouring traps thus 
created, so that $N_1+N_2=N$. Let  the difference in the condensate densities
be $z(0) =(N_{1}(0)-N_{2}(0))/N$. Let $\phi(0)$ be the initial {\it phase difference}
between the two condensates.  
We propose that in an actual experiment, immediately
after creating the laser sheet barrier, if the density difference and the phase 
difference between the condensates in the two traps  can be measured
as a function of time, then by substituting these 
experimentally measured functions in Eq. (3.17), the associated geometric 
phase $\phi_g$ can be  determined. As we have seen, $\phi_g$
will depend on system parameters as well as initial conditions. 

Theoretically, the evolution equations for $z(t)$ and $\phi(t)$ are given in Eqs. (2.8a) 
and Eq. (2.8b) respectively. The trap parameters are given in Eq. (2.9).
As an illustrative example, we have chosen the  parameters 
$\frac{\Lambda}{V}=5$ and $\Delta E=0$. This  corresponds to an interacting bose system 
in a symmetric trap.  With appropriate
initial conditions  $(z(0),\phi(0))$ for  both a librational orbit and and a
rotational orbit, we have solved Eqs. (2.8a) and (2.8b)
for $z(t)$ and $\phi(t)$ numerically. We then calculate the geometric phase $\phi_g$
for the above two types of orbits, by using Eq. (3.17).  
These are plotted in Figs. 8 and 9, respectively. As is obvious, 
the last point for $\phi_g$ on each of the plots, i.e., for the maximum value $t=T_{m}$, 
corresponds to  the geometric phase for a cyclic evolution, when the population 
difference and phase-difference between the two condensates evolve in such a way as
to return to their initial values after a time period $T_{m}$. 
All the other intermediate points correspond to non-cyclic evolutions. 
   
As should be clear, $\phi_g$ for other parameters $\Lambda$ and $\Delta E$
can also be computed by following our method, case by case. This would
enable one to study the variation of $\phi_g$ with trap parameters,
which would be useful in designing appropriate experiments
to measure it. 

 The possibility of another type of experiment to study tunneling between condensates
has been proposed by Williams {\it et al} \cite{Will}.  This hinges on the fact that it
has become possible to confine  a two-component bose condensate in the same trap, as follows.
Hall {\it et al} \cite{hall} first trapped and cooled $^{87}\rm{Rb}$  atoms in a 
magnetic trap in the $|f=1,m_{f}=-1>$ hyperfine state.  After condensation,
it is possible to populate the $|f=2, m_{f}=1>$ hyperfine state through 
a two-photon transition. In the presence of a weak magnetic field,
these states are separated in energy by $\omega_{o}$ (say).
Thus two different hyperfine states can exist
in the trap. A weak two-photon driving pulse  is applied which couples the two
states and consequently, atoms can get transferred  (or "tunnel") between 
the two condensates. In this model, it has been shown\cite{Will} that in the mean field
approximation, one obtains coupled equations for $z(t)$ and $\phi(t)$ 
almost identical in form to Eqs. (2.8), but with $\Lambda=0$ (i.e., non-interacting)
with the other parameters appropriately defined for the model, and hence all our 
results for the geometric phase are applicable here as well.  

Recently, Fuentes-Guirdi {\it et al} \cite{fuentes} have proposed a method
for generating a geometric phase in a coupled two-mode Bose Einstein
condensate, starting with a Hamiltonian for two condensates existing in
{\it different} hyperfine states. In addition to the experiments of Hall mentioned above,
condensates of  $^{87}\rm Rb$ atoms in hyperfine states $|f=1, m_{f}=1>$ 
and $|f=2,m_{f}=2>$  have been  produced  experimentally\cite{myatt}. 
Likewise, condensates of $^{23}\rm Na$ atoms with
$|f=1, m_{f}=1>$ and $|f=1, m_{f}=0>$   have also
 been created\cite{meisner}. Using the Schwinger angular momentum
representation, the Hamiltonian  describing  
two coupled hyperfine states $|A>$ and $|B>$ can be expressed 
in the form \cite{fuentes}
$$
H_{hf}=\alpha_{0}J_{z}+\beta_{0}J_{z}^{2}+\gamma_{0}[J_{x}\cos\phi_{D}+
J_{y}\sin\phi_{D}],
\eqno(6.1)$$
Here, $(J_x,J_y,J_z)$  are the components of an effective
`mesoscopic' spin ${\bf J}$, since it can be shown that $J$ is proportional
to  the total number of atoms  $N$  in the condensate, which is of 
the order of $10^{4}$. In Eq. (6.1), 
$\phi_{D}=D~t$, $D$ being the detuning frequency of the laser
which couples the two hyperfine states. Further,
$$
\alpha_{0}=(\omega_A-\omega_B)+(2J-1)(U_A-U_B),
\beta_{0}=\frac{U_A+U_B-U_{AB}}{2},
\eqno(6.2)$$
and $\gamma_{0}$ is the strength of the laser-induced drive term 
that couples the two levels. 

Interestingly, if we write the components of ${\bf J}$ in the form 
$$
{\bf J}=(J_x,J_y,J_z)=J(\sin \alpha \cos\phi, \sin\alpha\sin\phi,
\cos \alpha)
$$
in Eq. (6.1), then on setting $\phi_D=0$, 
$H_{hf}/J$  becomes identical to our Hamiltonian in Eq. (2.10),
on identifying  $\alpha_{0}=\Delta E$,  $\beta_{0}=\Lambda/2$ and 
$\gamma_{0}=-V$. Conversely, 
if an external driving field  phase $\phi_D$ is subtracted from $\phi$ 
in Eq. (2.10), we would essentially obtain Eq. (6.1). Thus our results
for the geometric phase will be valid for that case too, with the appropriate  
parameters substituted.

However,  we point out that in the setting of \cite{fuentes}, one varies
the BJJ system parameters $\alpha_{0}$ and $\beta_{0}$ adiabatically to produce 
a closed circuit in parameter space,i.e., one considers an adiabatic, 
cyclic evolution in parameter space, with an associated geometric phase. 
In contrast, in our setting,
we have fixed system parameters, for which  we obtain the geometric phase 
for  both cyclic and noncyclic (and nonadiabatic) evolutions 
on the unit sphere, i.e., in "spin" space. In other words, as the population difference
and phase difference between the two hyperfine states evolve in time, the system
follows a  path in this projective space, and as we have shown, an associated geometric phase
can be defined.       

Experimental techniques to produce two condensates in close
proximity has been suggested  recently by Chikkatur {\it et al}\cite{chik}
in the context of a technique to produce a continuous source of a Bose-Einstein condensate.
It would  be interesting to  study the tunneling between the condensates in such a set up
 if feasible, i.e., investigate the evolution of population differences and phase 
differences, to find the associated geometric phase.

Geometric phases have been recently shown to have relevance in the implementation of 
fault-tolerant  quantum computation \cite{zana}, and in the creation of vortices 
in a condensate \cite{petros}. We hope that our results will have
applications in these contexts as well.

\newpage

\begin{figure}[h]
\includegraphics{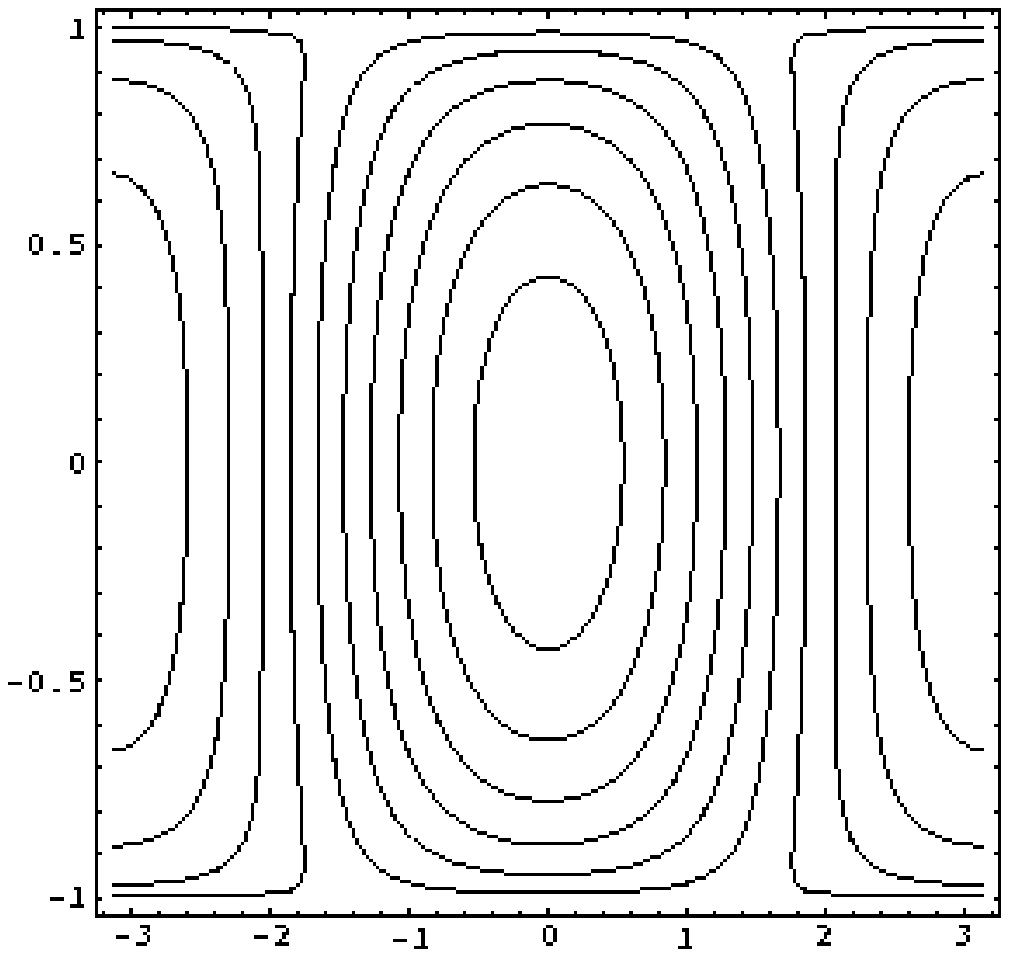}
\caption {\label{fig1}
Phase portrait of BJJ evolution (Eq. (2.13)) for an interacting bose system
  with $\Lambda=0.5$, in a symmetric trap.  
 }
\end{figure} 

\begin{picture}(100,200)(-100,-10)
\put(125,250) {\large $\phi$}
\put(-40,400) {Z}
\end{picture} 

\newpage

\begin{figure}[h]
\includegraphics{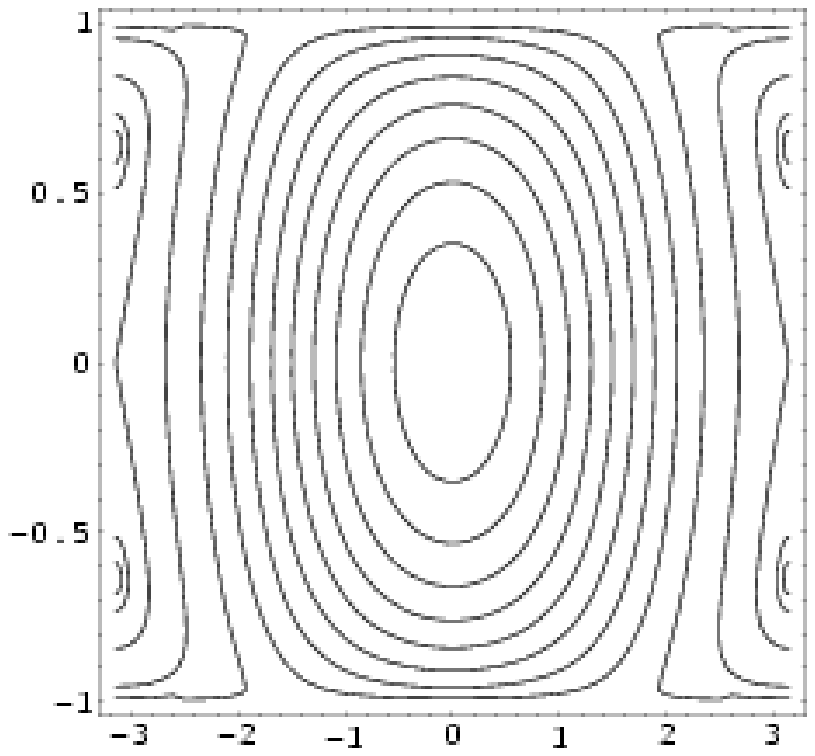}
\caption {\label{fig2}
Phase portrait of BJJ evolution (Eq. (2.13)) for an interacting bose system 
  with $\Lambda=1.3$, in a symmetric trap. The trapped states at $\phi = \pi$
  are clearly visible here.   
 }
\end{figure}  

\begin{picture}(100,200) (-100,-10)
\put(-40,400) {\large Z}
\put(125,250) {\large $\phi$}
\end{picture}

\newpage

\begin{figure}[h]
\includegraphics{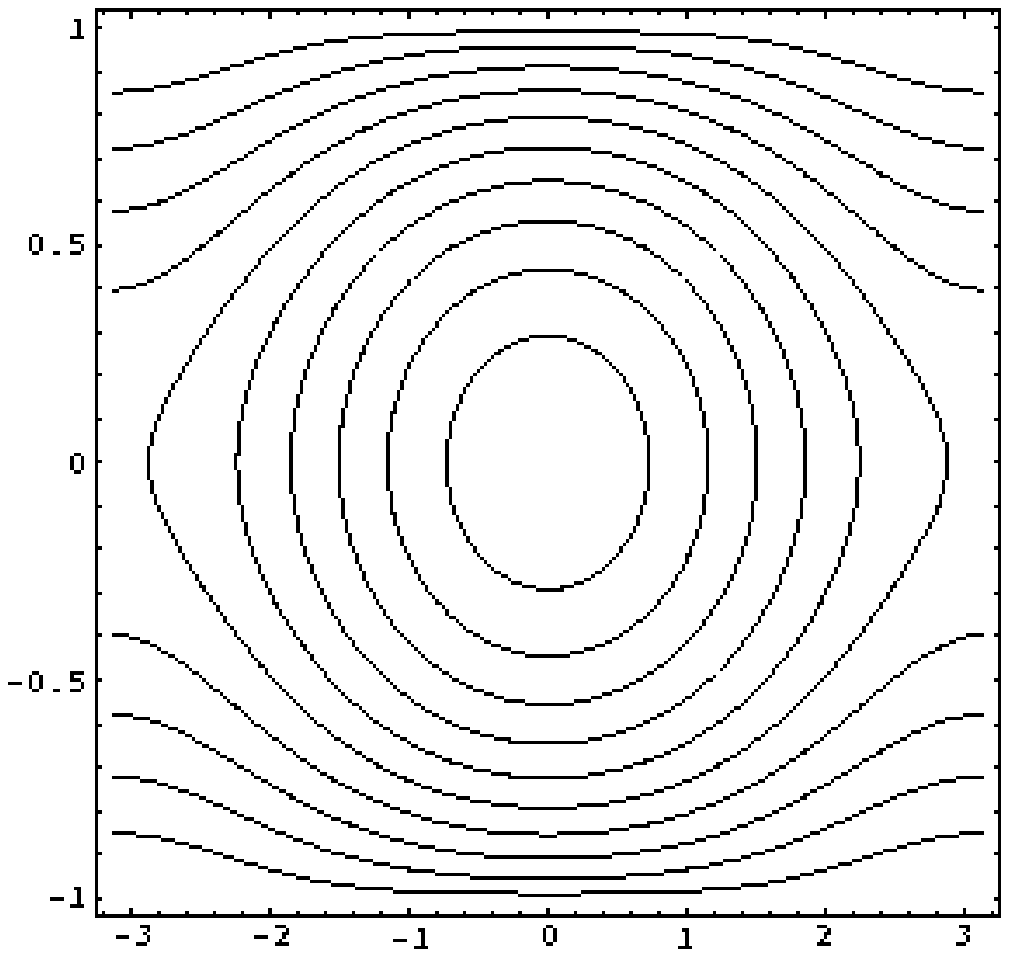}
\caption {\label{fig3}
Phase portrait of BJJ evolution (Eq. (2.13)) for an interacting bose system 
  with $\Lambda=5$, in a symmetric trap.  
 }
\end{figure}  

\begin{picture}(100,200)(-100,-10)
\put(-40,400) {\large Z}
\put(125,250) {\large $\phi$}
\end{picture}

\newpage

\begin{figure}[h]
\includegraphics{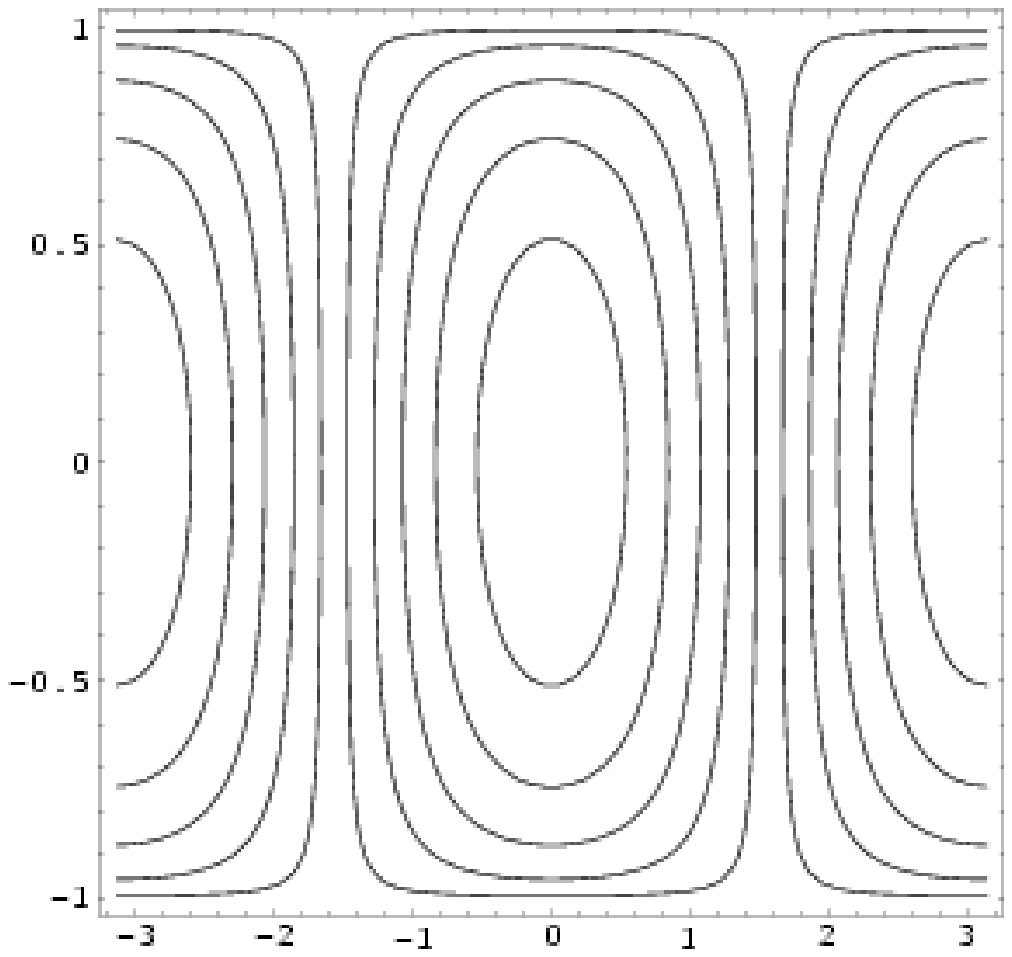}
\caption {\label{fig4}
Phase portrait of BJJ evolution (Eq. (2.14)) for a non-interacting bose system 
  in a symmetric trap with $\Delta E=0$  
 }
\end{figure}  

\begin{picture}(100,200)(-100,-10)
\put(-40,400) {\large Z}
\put(125,250) {\large $\phi$}
\end{picture}

\newpage

\begin{figure}[h]
\includegraphics{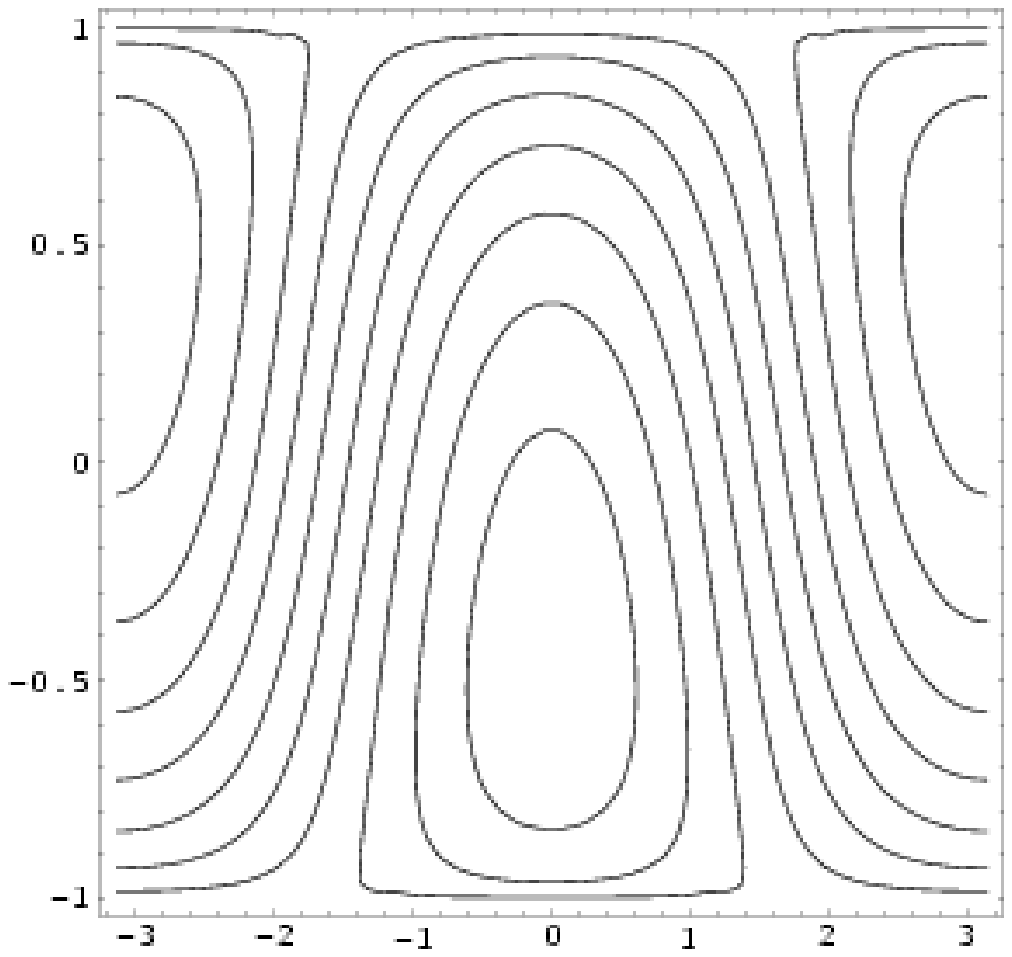}
\caption {\label{fig5}
Phase portrait of BJJ evolution (Eq. (2.14)) for a  non-interacting bose system
 in an asymmetric trap with $\Delta E=0.5$  
 }
\end{figure}  

\begin{picture}(100,200)(-100,-10)
\put(-40,400) {\large Z}
\put(125,250) {\large $\phi$}
\end{picture}

\newpage

\begin{figure}[h]
\includegraphics{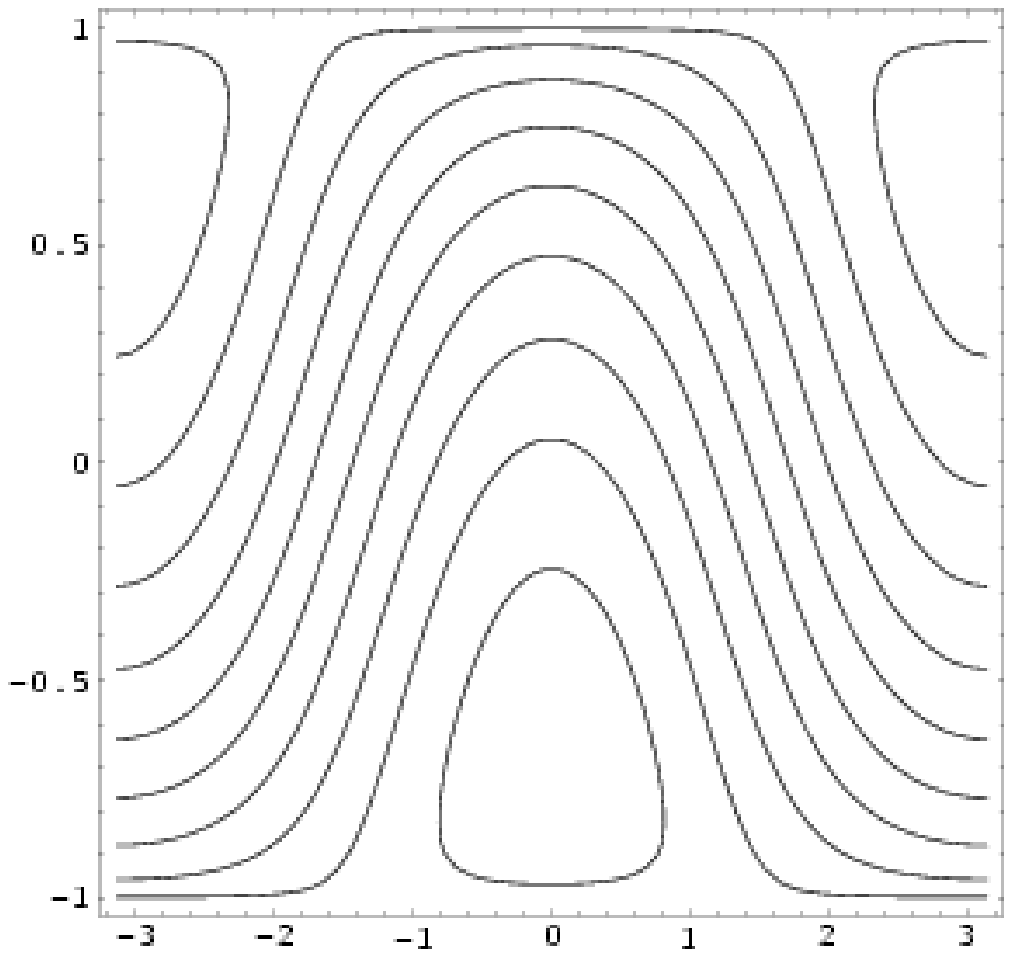}
\caption {\label{fig6}
Phase portrait of BJJ evolution (Eq. (2.14)) for a non-interacting bose system 
  in an aymmetric trap with $\Delta E=1.0$  
 }
\end{figure}  

\begin{picture}(100,200)(-100,-10)
\put(-40,400) {\large Z}
\put(125,250) {\large $\phi$}
\end{picture}

\newpage  

\begin{figure}[h]
\includegraphics{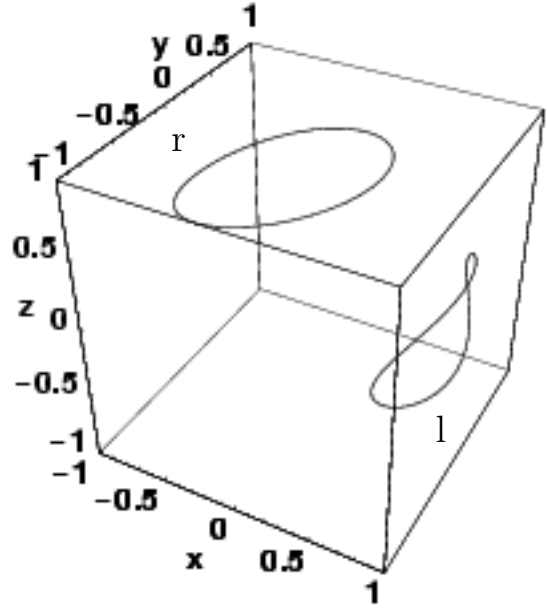}   
\caption {\label{fig7}
BJJ evolution of the unit vector {\bf r}
  (see Eq. (4.2)) on the unit sphere: Paths  corresponding
 to a librational orbit and a rotational orbit (labeled r
 and l respectively in the plot) in 
the phase space portrait of the BJJ Hamiltonian for  a
 symmetric trap with $\Lambda = 5$ (see Fig. 2)are shown.}
 \end{figure}  

\begin{picture}(100,200)(-100,-10)
\put(80,460) {\large r}
\put(180,350) {\large l}
\end{picture} 

\newpage

\begin{figure}[h]
\includegraphics{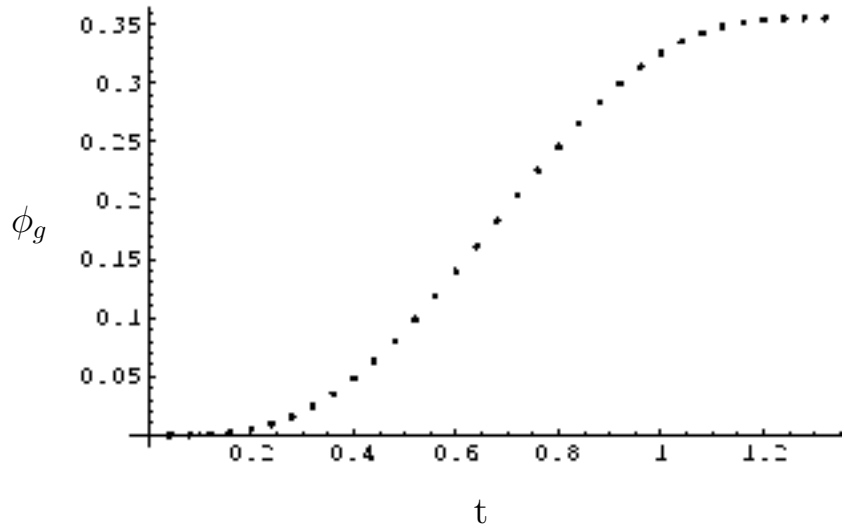}
\vskip .5cm
\caption {\label{fig8}
Evolution of the geometric phase as a function of time
 over a period for a librational orbit (oscillation about
 the zero-state) at $\Lambda = 5$, with initial conditions
$(z,\phi) = (.3,0)$ corresponding to orbit l in Fig. 7} 
\end{figure}

\begin{picture}(100,200)(-100,-10)
\put(-50,390) {\large $\phi_g$}
\put(125,280) {\large t}
\end{picture}

\newpage

\begin{figure}[h]
\includegraphics{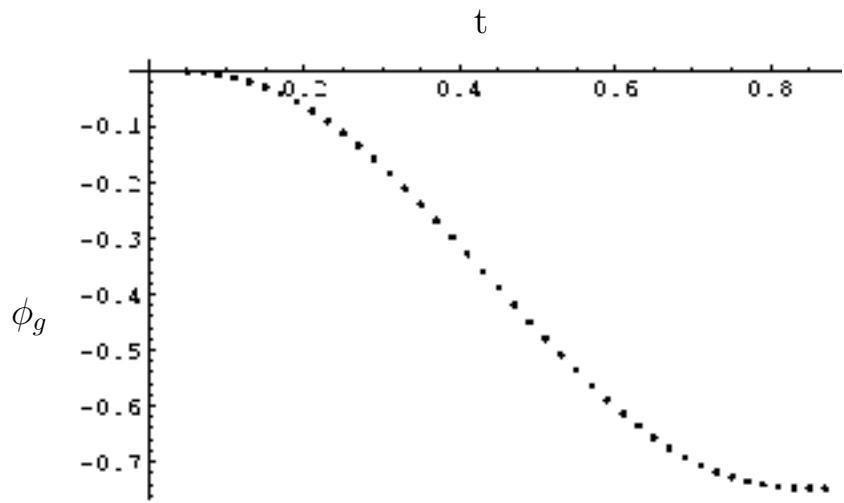}
\vskip .5cm
\caption{\label{fig9}
Evolution of the geometric phase as a function of time
 over a period for a rotational orbit at $\Lambda = 5$ with initial
conditions $(z,\phi) = (.9,0)$ corresponding to orbit r in Fig. 7.  
 }
\end{figure}

\begin{picture}(100,200)(-100,-10)
\put(-50,350) {\large $\phi_g$}
\put(125,460) {\large t}
\end{picture}

\end{document}